\documentclass[twocolumn]{article}
\usepackage[utf8]{inputenc}
\usepackage{authblk}
\usepackage{url}
\usepackage[english]{babel}
\usepackage[ruled,vlined]{algorithm2e}
\usepackage[a4paper, margin = 0.8in]{geometry}
\usepackage{graphicx}
\title{ Graph500 from OCaml-Multicore Perspective}
\author{Shubhendra Pal Singhal}
\date{}
\usepackage{listings}
\begin{document}

\maketitle

\section*{Abstract}
OCaml is an industrial-strength, multi-paradigm programming language, widely used in industry and academia. OCaml was developed for solving numerical and scientific problems involving large scale data-intensive operations and one such classic application set is Graph Algorithms, which are a core part of most analytics workloads. In this paper, we aim to implement the graph benchmarks along with the performance analysis. Graph500 is one such serious benchmark which aims at developing data intensive applications requiring extreme computational power. We try to implement Graph Construction, BFS, Shortest-Path problems using the desired specifications and rules posed by graph500. This paper aims at providing a clear direction of choices of several data structures used, algorithms developed and pose a reason behind every step of program. The first few sections of the paper discusses a formal approach to the problem with a small guide for starters in OCaml. The latter sections describe the algorithms in detail with the possibilities of future exploration and several mistakes which we committed or encountered whilst approaching the solution. All performance metrics were tested on  Intel(R) Xeon(R) Gold 5120 CPU @ 2.20GHz 24 core machine. Every section talks about the initial performance failures encountered, which will help analyse and prioritise our preferred implementation from a performance perspective.   \newline \newline
\textit{\textbf{Additional Keywords} : Graph500, BFS, SSSP, Kronecker, Sparse Matrix, CSR}

\section{Introduction}
This section explores some of the basic fundamentals necessary for OCaml and functional programming in general.
\subsection{Diving into Functional Programming : OCaml}
Imperative languages like C++, C, Python are based on a fact that they focus on how to execute (by defining each step of flow) and define the control statements to determine the change of the state. Whereas in functional programming, it focuses on the program logic rather than the flow of the execution.\newline

To begin with OCaml, there are a few things which are essential to follow\cite{1} : 
\begin{itemize}
    \item  \textbf{Immutable} : In imperative languages, we declare and change the value of a variable as (in C++) : 
    \begin{lstlisting}{language = C++}
    int val = 3;
    val = 4;
    \end{lstlisting}
    In this case, this variable was mutable, i.e. the value of the variable could be changed. But in OCaml, every variable is immutable i.e. once a variable is defined and declared, it can never be changed.
    
    \item \textbf{Recursion} : Now, if you realise that mutablility is not preferred in the functional programming language, loops would also not be preferred in this same environment. So every function in OCaml is defined in a recursive manner rather than using loops.
    Illustration (Nth($>$3) Fibonacci number) given below would make it clear :
    \newline 
    
    \textit{Imperative} :

    \begin{lstlisting}
    function fibonacci (n) : 
        fib(0) = 0
        fib(1) = 1
        for i = 2 to n-1 : 
            fib(i) = fib(i-1) + 
                     fib(i-2)
        return fib(n-1)
    \end{lstlisting}

    \textit{Functional} : \newline 
    \begin{lstlisting}
    let rec fibonacci a b index n = 
        if index = n-1 then a+b
    else 
        fibonacci (a+b) (a) (index+1) n
        
    let () = 
        let answer = fibonacci 0 1 2 n
    \end{lstlisting}
    In this case, you would realise how recursion is adopted as a primary concept of constructing the logic in OCaml.
    
    \item \textbf{Tail Recursion} : 
        Notice the difference between these two factorial functions
        \begin{lstlisting}
    let rec fact n = 
        if n > 1 then n*fact (n-1)
    else 1 
    \end{lstlisting}
    whereas,
    \begin{lstlisting}
    let rec fact n ans = 
        if n = 1 then ans
    else fact (n-1) (ans*n)
    \end{lstlisting}
    \begin{figure}
    \centering
        \includegraphics[scale = 0.07]{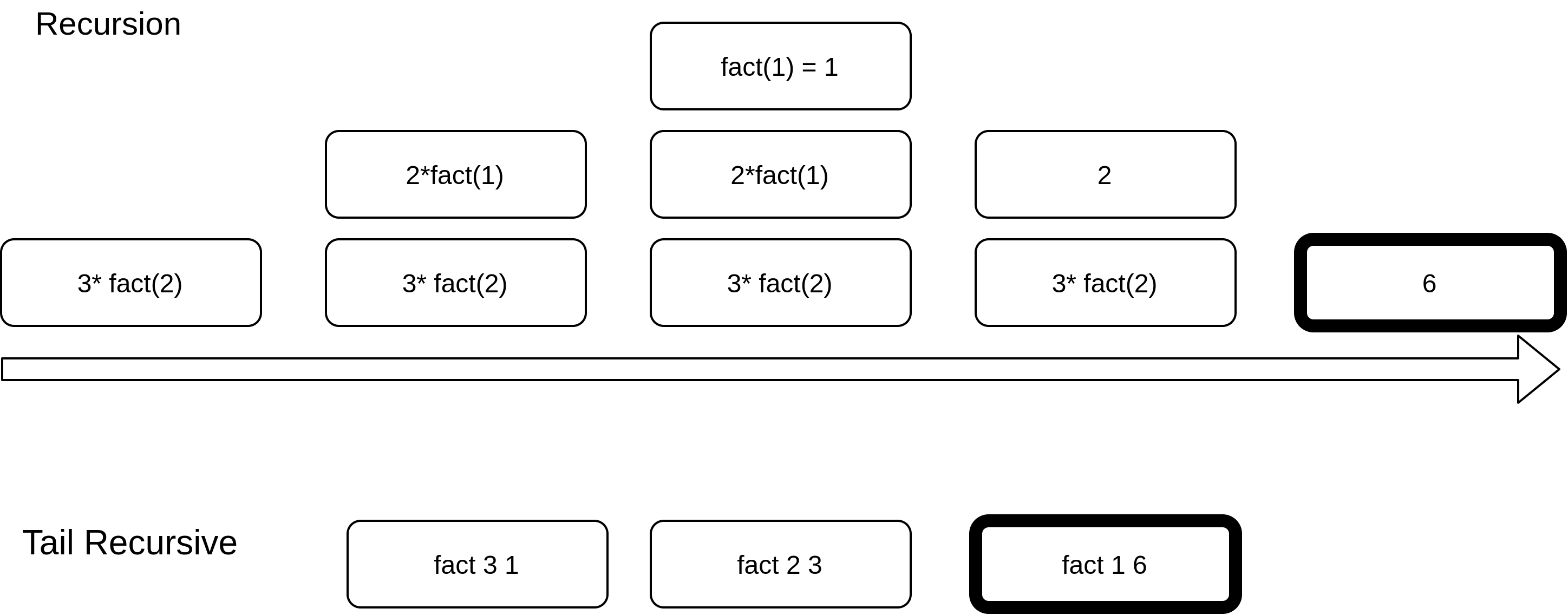}
        \caption{Recursive vs Tail Recursion}
        \label{tail}
    \end{figure}
    In the first function, the recursion has to allocate the memory for every stack frame for returning the factorial of a number whereas the second function allocates the size of only one stack frame, avoiding the problems of stack overflow\cite{2} in large benchmark algorithms, illustrated in Fig.\ref{tail}.
\end{itemize}
Some of these idiomatic representations in OCaml are peculiar for functional programming and must be kept in mind while developing the logic of the program\cite{2}.

\subsection{Experimenting with OCaml}
There are various ways of understanding OCaml system as a whole. One of the ways which proved to be effective is to : \newline
\begin{itemize}
    \item Test various small programs like constructing arrays, lists, factorial, fibonacci, hashmaps\cite{3}.
    \item After you are comfortable debugging the errors related to these basic syntax, try building a calculator with float and int operations.
    \item Next, comes the if else operations with more than one statement in every condition. Try understanding the various errors related to ';', begin..end statements. Errors would not be very clear in these scenarios so start with smaller programs.
    \item Try solving atleast 40-50 "99 OCaml problems"\cite{4} and test them on utop.
    \item Final and the last step is to make a graph algorithm like Shortest Path problem from a vertex . After testing it on utop, try compiling with ocamlopt, ocamlc(not very efficient) and dune. Try measuring time of these executables to realise the inner structure of OCaml using profiling from perf.
\end{itemize}

\section{Graph500}
Graph500\cite{5} is a large scale graph benchmark (“Search” and “Shortest-Path”), which intents to develop a compact application that has multiple analysis techniques (multiple kernels) accessing a single data structure representing a \textbf{weighted, undirected graph}. In addition to a kernel to construct the graph from the input tuple list, there are two additional computational kernels to operate on the graph.
\newline
This benchmark includes a scalable data generator which produces edge tuples containing the start vertex and end vertex for each edge. The first kernel constructs an undirected graph in a format usable by all subsequent kernels. \textit{No subsequent modifications are permitted to benefit specific kernels}. The second kernel performs a breadth-first search of the graph. The third kernel performs multiple single-source shortest path computations on the graph\cite{5}.
\begin{algorithm}
\SetAlgoLined
1. Generate the edge list. \\
2. Kernel 1 : Construct a graph from the edge list.\\
3. Kernel 2 : Compute the parent array and bfs tree.\\
4. Kernel 3 : Compute the parent array and the distance array.\\
5. Compute the performance numbers.
 \caption{Basic Outline of Graph500}
\end{algorithm}

\subsection{Kronecker : Generating edge list}
The kronecker algorithm takes two inputs, scale and edgefactor which are basically : \newline
Graph $G = (V,E)$ where,
\\$\#V$ = $2^{scale}$ and,\\ $\#E = \#V *$ edgefactor.
\newline
The graph generator creates a small number of multiple edges between two vertices as well as self-loops. Multiple edges, self-loops, and isolated vertices may be ignored in the subsequent kernels but must be included in the edge list provided to the first kernel. As a final step, vertex numbers must be randomly permuted, and then the edge tuples randomly shuffled\cite{5}.
\newline
It generates the graph in the form of a matrix of dimension $3\times \#E$, where every column denotes (start-vertex, end-vertex, weight) tuple.\newline
Main challenge behind the implementation of kronecker is the choice of data structure and its linking to all the kernels.\newline
Generating the random vertex set in kronecker requires only one consideration i.e. vertices produced from the initiator probabilities should be an array instead of a list as mentioned in the original algorithm. This would boost up the \textbf{execution time by almost 3x for a uni-core system, when tested for a toy problem (1.4MB) graph containing 40,000 edges}, which is a really huge improvement provided there is just a single change. 
\newline
Regarding the linking of kronecker to other kernels, we have chosen to store the graph in a text file in hard drive named as "kronecker.txt". This is important as if we choose to run kronecker from kernels individually, random permutation being used may not result in the same graph generation. So to avoid any errors, kronecker runs separately. Also, there is no need to parallelise kronecker as this is not being timed or considered for performance benchmarking. 

\subsection{Kernel 1 : Construction of Graph}
There are various ways of representing graphs\cite{8}, adjacency matrix, lists, hashmaps, CSR/Yale format etc. Analysing all of these individual representations : \newline
\begin{itemize}
    \item \textbf{Adjacency Matrix} : Graph500 consists of parameters - scale and edgefactor where number of edges $\#E$ are far more less than the case where every pair of vertex $(u,v)$ has a edge (consider any parameter setting, (26,16),(42,16) etc.) i.e. \newline \newline
    $\#E << \#V*(\#V - 1)/2  $, \newline
    which indicates that there would be a lot of zeroes in adjacency matrix thereby occupying a lot of unnecessary space and traversal time.\newline
    
    \item Next question that arises is, that construction of graph needs to be efficient from the bfs and shortest path's point of view. BFS traversal would atleast visit every node once implying that in the case of adjacency list, the key head would have to be searched for in-order to retrieve its neighbouring vertex. So in order to reduce such complexity, we came to the conclusion of using Hashmaps, where every head is a key and every key contains the list of its neighbouring vertices.  
\end{itemize}
After implementing the kernel2 and kernel3 using Hashmap as a concept, we measured the profiling data of the kernels using perf. This was a big step taken in deciding whether our implementation is efficient or not. Eventually, we figured that in case of hashmaps, garbage collector\cite{1} has to revisit the whole hashmap in order to maintain its allocation and secondly the hash function might not necessarily reduce the lookup time for every graph.\newline
Therefore, we looked at the other very efficient implementation known as CSR format where the graph is stored in three arrays. \newline
\begin{figure}
    \centering
    \includegraphics[scale = 0.09]{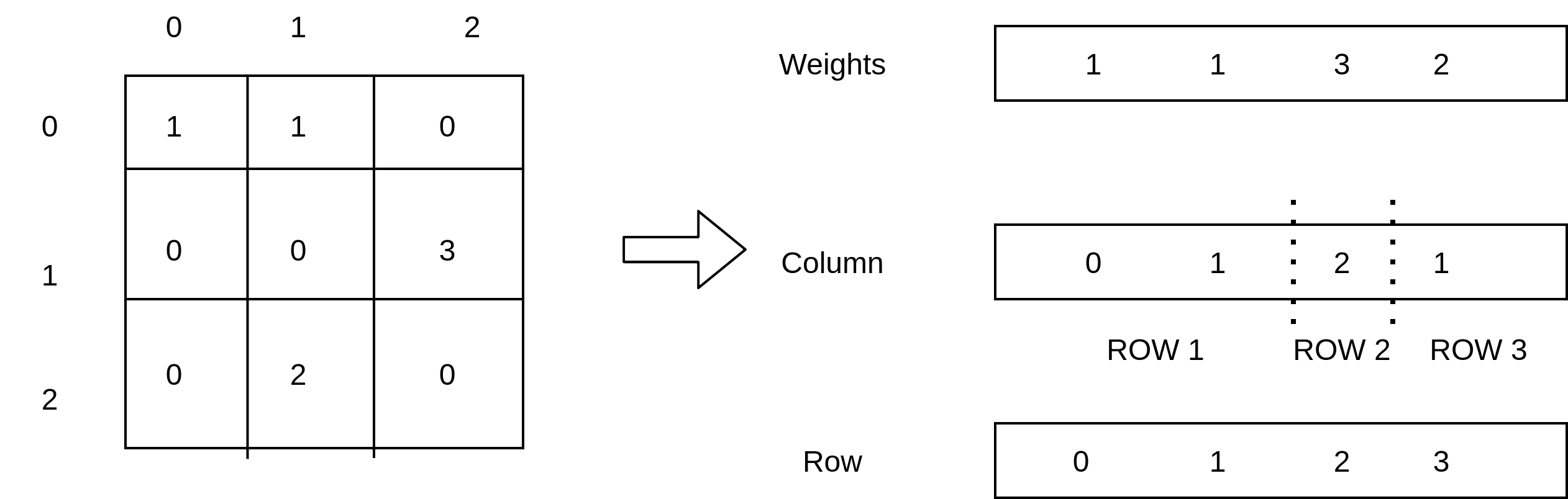}
    \caption{CSR Format}
    \label{csr}
\end{figure}
CSR format is illustrated in Fig.\ref{csr} where there are a few key points to note for the implementation. There would be cases where the whole row is empty. In such a case, we need to repeat the same number as the previous row index. Another important point is that we need to store the last row's index too instead of leaving it by calculating it from subtracting the last index with the array's length because there might be a case where number of rows are n in number, but last few rows are empty. \newline
The CSR format, when implemented and profiled led to a conclusion that while representing sparse matrix, GC performed well but despite the access being O(1), it performed worse by almost 5 to 10 seconds on an average for same core and graph size, than a Hashmap.\newline

But one important thing to note here was that in Hashmap we didn't require sorting of kronecker array whereas in CSR format we require the sorted array and sorting $\sim$ million elements will negate/overpower the speedup gained from the O(1) access. \textit{Both of the implementations have their own pros and cons but when it comes to performance measurement both performs equally worse}. Just because GC is maintained well, usage of CSR is not justified.\newline
Eventually, the main aim of implementing kernel1 is to parallelise it. So, in case of the hashmaps, formation of the key-list can be parallelised by dividing the array tuple generated by kronecker squarely for the threads to work on different segments of that array. So for instance there are two threads working on array, one being from 0 to $4^{th}$ index and other from $5^{th}$ to $9^{th}$. In this case, construction of key-list pairs have to take in consideration that in OCaml, Hashmap module is not thread safe implying that if two threads access the same hashmap at the same time, they might not write the data in correctly or might even miss out. Therefore, we have used Lockfree Hashmap\cite{6} in such case, where the key-list i.e. write operation happens sequentially, and in this case, binding of a key is replaced by a new list every time an element is added to the list. As the array is not sorted, so possibility of starvation for threads is not possible as the thread will not acquire the hashmap entirely for one key because of its randomness(key-list binding does not happen in sorted order of keys). In this scenario, \textbf{an average speedup over all input cases of only 3.3x could be witnessed on 8, 12 and 16 cores}. As the size of the graphs increases(in gigabytes), the speedup increases(upto 4x) relatively with more number of cores. \newline
The parallel implementation of kernel1 using CSR format required an additional processing, because the division might result in rows being processed in two different threads because of which a maximum function had to be included. For instance if thread 1 says that row 1 is upto $4^{th}$ index whereas thread 2 says that row 1 is upto $8^{th}$ index, its a possibility that both the threads are processing the same row. A mini-prototype of such (without optimising the storage space) was tested out in-order to visualise whether we receive any speedup or not, but turns out the speedup obtained after implementing a small base case was just 1.52x, which overall leads to a decrease in performance as compared to hashmap.  \newline
\textit{The choice of data structure for graphs is still a big question mark which needs to be addressed in-order to obtain a good record-breaking speedup}. The following kernels have both been described from CSR and Hashmap perspective in the form of general algorithm for reference as the only thing which differs in their usage is \textit{accessing the list of neighbouring nodes for a particular node}.

\subsection{Kernel2 : BFS}
\subsubsection{Sequential BFS}
Sequential BFS implemented using a queue is described below. BFS tree can be constructed from a level array where level of every vertex is the depth when represented in the form of a tree. The algorithm is described in Algorithm \ref{BFS}.

\begin{algorithm}
\SetAlgoLined
\label{BFS}
\caption{Sequential BFS}
Let $G = (V,E)$\\
visited.(index) $\leftarrow$ 0 $\forall$ index
$\epsilon$ $E$ \\
level.(index) $\leftarrow$ (-1) $\forall$ index
$\epsilon$ $E$ \\
visited.(root) $\leftarrow$ 1, 
level.(root) $\leftarrow$ 0\\
Enqueue($Q$, root)\\
\SetKwFunction{FMain}{BFS}
\SetKwProg{Fn}{Function}{:}{}
\Fn{\FMain{$Q$}}{
\While{$Q \neq \phi$}{node = Dequeue($Q$)\\
        \For{each $u$ $\epsilon$  Adj[node]}{
            \If{visited.($u$) $\rightarrow$ 0}{
                level.($u$) $\leftarrow$ level.(node) + 1\\
                Enqueue($Q$, $u$)\\
                }}
        visited.(node) $\leftarrow$ 1} 
        \KwRet level
}
\end{algorithm}

\subsubsection{Parallel BFS}
\begin{figure}
    \centering
    \includegraphics[scale = 0.09]{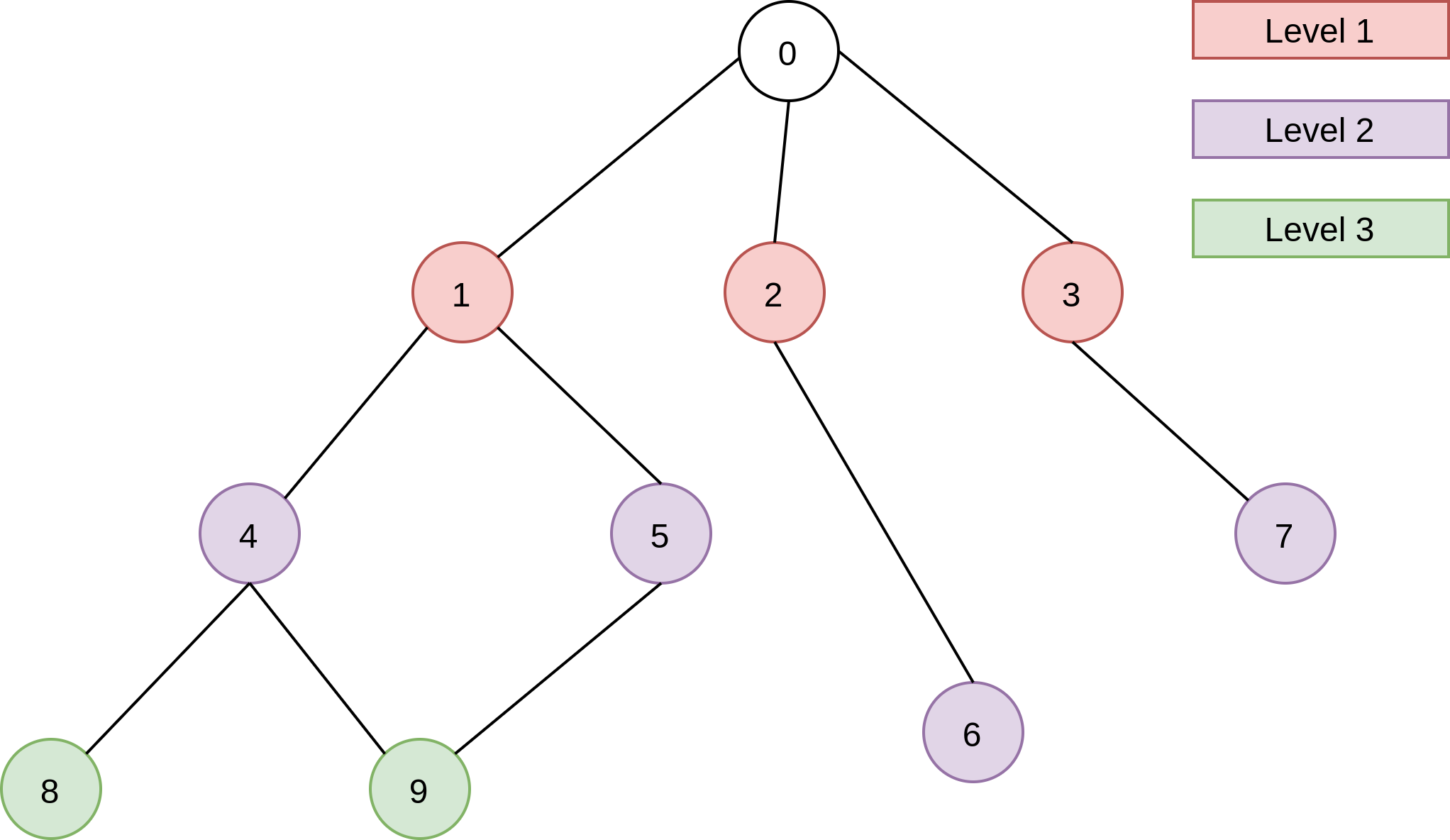}
    \caption{Levels in BFS}
    \label{level}
\end{figure}

\begin{figure}
        \centering
        \includegraphics[scale = 0.09]{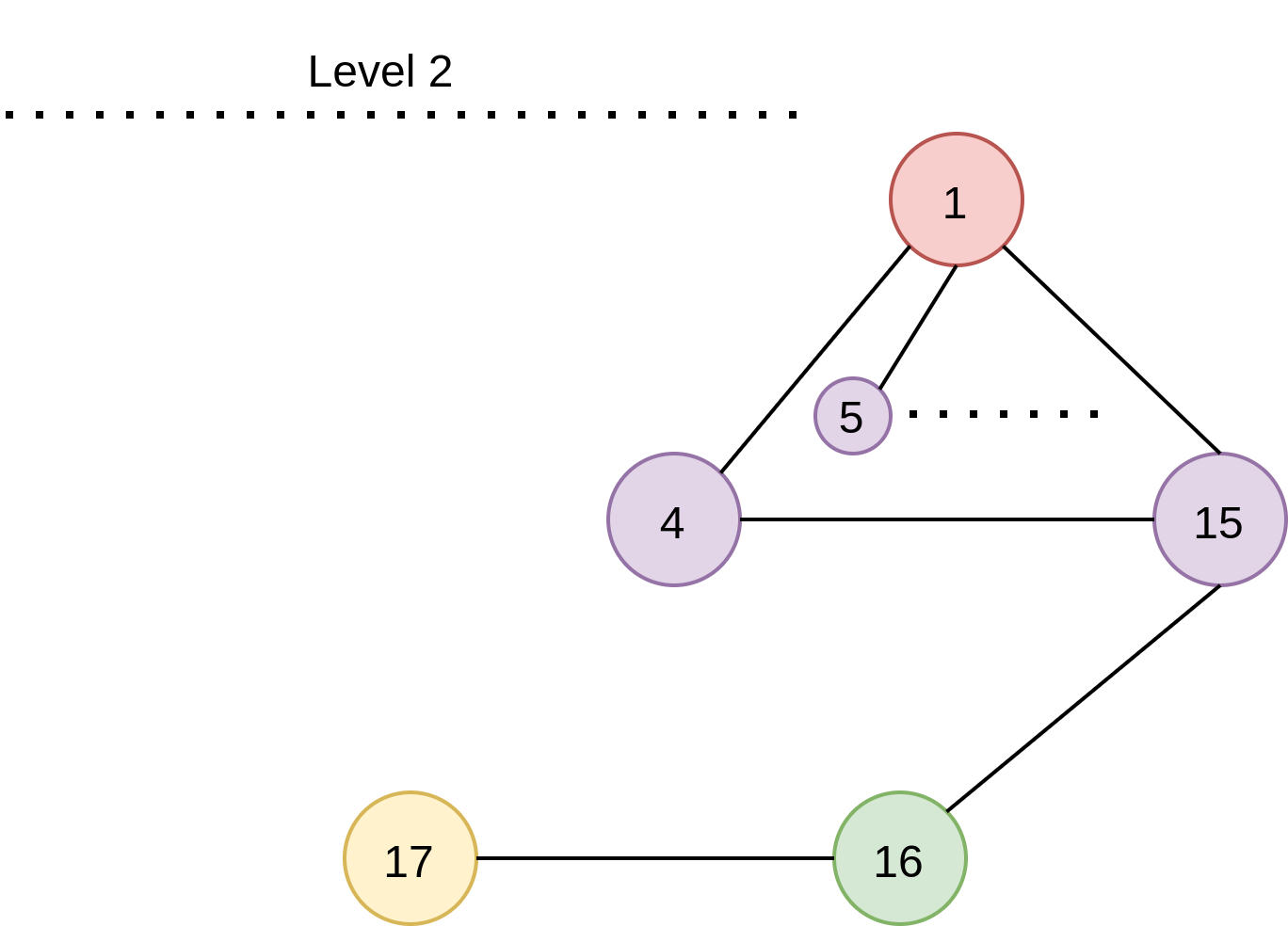}
        \caption{Failure case}
        \label{fail1}
    \end{figure}

Using the basic idea of level synchronization, BFS can be parallelised. Level synchronization\cite{7} means that for a root node, threads can work in parallel for a particular level, and after all threads have completed their work for all the neighbouring nodes and the level is complete, it will allow the second iteration of the loop to be executed. This can be explained easily with the help of a figure illustrated in Fig. \ref{level}.\newline
After understanding what the level synchronization is, now its the time to understand that this synchronization was explained for a node at a level. But we can operate on all the nodes parallely on a level. So therefore, comes the another concept known as "Process an entire level in one step in parallel"\cite{7}.\newline
\textit{Importance of Level Synchronization} :\newline
There is a node connected to two nodes of levels 2 and 3. Now if level synchronization is not ensured, the thread operating from node having level 3 might update it to 4, before it has been updated to 3 by the other thread. One argument which can be posed is that the minimum of the indices (except -1) will be maintained using a lock.
    If the lock is not maintained, there might be an instance where both the threads read the same value -1, and one overwrites the other because both their value's were minimum in point of view of threads. But, even if we maintain a lock, here comes another failure scenario. \newline
    As illustrated in Fig\ref{fail1}, imagine there is a reasonable time difference between the updation of level of the node from 4 to 3. In this period of time, if that node is further connected to other nodes, the level for those nodes is wrongly updated and if these nodes are not connected to any other nodes, then they will be not be revisited by any other thread. And if we argue that after updation, we will update all the subsequent affected nodes again, it is going to be sheer waste of time as the time taken would then be directly proportional to the number of updations for other nodes which might be huge if the vertex is densely populated. Note that its the vertex's locality being referred to as dense not the graph's.\newline 

Understanding both the concepts mentioned above, carefully examine the algorithm described in Algorithm \ref{PBFS}.
\begin{algorithm}
\SetAlgoLined
\label{PBFS}
\caption{Parallel BFS}
Let $G = (V,E)$\\
level.(index) $\leftarrow$ (-1) $\forall$ index
$\epsilon$ $E$ \\
$l$ $\leftarrow$ 0 \\
level.(root) $\leftarrow$ 0\\
Enqueue($Q$, root)\\
\SetKwFunction{FMain}{BFS}
\SetKwProg{Fn}{Function}{:}{}
\Fn{\FMain{$Q$}}{
\While{$Q \neq \phi$}{$Q' \leftarrow \phi $\\ \For {each node in $Q$ in parallel}{
        \For{each $u$ $\epsilon$  Adj[node] in parallel}{
            \If{level.($u$) $\rightarrow$ -1}{
                level.($u$) $\leftarrow$ $l$ + 1\\
                Enqueue($Q'$, $u$)
                }}}
                $Q \leftarrow Q'$ \\ 
                $l$ $\leftarrow$ $l$ + 1} 
        \KwRet level
}
\end{algorithm}
In parallel bfs, in-order to maintain the barrier of level synchronization and constraint on the number of threads, we have implemented it by first accumulating the neighbouring nodes of all the nodes in a level into a list parallely and then update the levels of all these elements parallely in order to maintain the synchronization. Referring to threads required in first parallel loop as loop1 and second as loop2, in this way we require the number of thread creations equal to maximum of (loop1, loop2) whereas the actual algorithm demands loop1 $\times$ loop2. Although we can re-use the threads after they complete their work, but we cant rely on this fact initially while creating the number of threads.\newline

\subsubsection{Performance Metrics}
\begin{figure}[h]
    \centering
    \includegraphics[scale = 0.3]{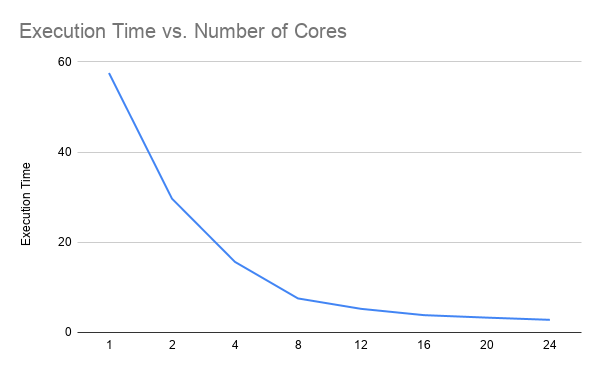}
    \caption{Execution Time vs Number of Cores}
    \label{exec2}
\end{figure}

\begin{figure}[ht]
    \centering
    \includegraphics[scale = 0.3]{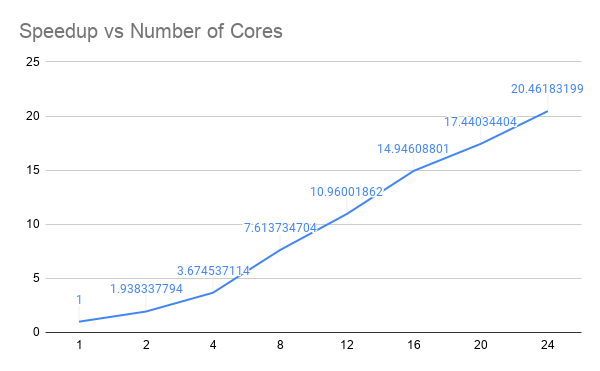}
    \caption{Speedup vs Number of Cores}
    \label{speedup2}
\end{figure}

A massive speedup reaching a maximum of \textbf{20x on Intel(R) Xeon(R) Gold 5120 CPU @ 2.20GHz on 24 cores} is obtained for a sample amongst the many test cases, of a graph of vertices $\sim$ 4000 and edges $\sim$ 1.2 lac, which has been shown in the Fig.\ref{exec2} and Fig.\ref{speedup2}. For bigger graphs, with \textbf{edges $\sim$ 3 lac, the speedup reaches upto 27.8x}. With higher sets as mentioned in graph500 with (scale, edgefactor) = (46,16), a really impressive speedup can be witnessed. Due to the limitation of the machine and the kronecker output, we could test it for medium sized graphs. (Graph500 assumes such graphs as toy sets.)

\subsection{Kernel 3 : Single Source Shortest Path}
\subsubsection{Sequential SSSP}
Sequential SSSP is performed by returning distance and parent arrays for every node as described in Algorithm \ref{SSSP}.
\begin{algorithm}
\SetAlgoLined
\caption{Sequential SSSP}
\label{SSSP}
\SetKwFunction{FMain}{Dijkstra}
\SetKwProg{Fn}{Function}{:}{}
\Fn{\FMain{G,S}}{
 Let $G = (V,E)$ and
 Q be the vertex set \\
 \For {every $v$ $\epsilon$ $Q$}{            
        dist[$v$] ← $\infty$ \\                 
        prev[$v$] ← -1 \\ 
        $Q$ $\leftarrow$ $Q$ $\cup$ \{$v$\}}               
dist[S] ← 0 \\                      
\While {$Q$ $\neq$ $\phi$}{
    $u$ $\leftarrow$ vertex in $Q$ with min dist[$u$]\\         
    \For {each $v$ $\epsilon$ $Q$ and $v$ $\epsilon$  Adj[$u$]}{
            alt $\leftarrow$ dist[$u$] + weight($u, v)$\\
              \If {alt $<$ dist[$v$]}{              
                  dist[$v$] $\leftarrow$ alt \\
                  prev[$v$] $\leftarrow$ $u$
                    }
                }
            }
      \KwRet dist, prev
}
\end{algorithm}

\section{Acknowledgments}
This paper and the research behind it would not have been possible without the exceptional support of my supervisor, KC Sivaramakrishnan and organisation OCaml Labs, UK. We are thankful for the generous grant offered by Tezos Blockchain.  

\section{Future Work}
\subsection{Parallel SSSP}
There are various algorithms like delta stepping, radius stepping which are quite popular. Looking into a possibility of just parallelising a for loop in SSSP which explores all the neighbours of a vertex, wouldn't yield you good parallelism. We tried this sort of parallelism and ended up with only 2.1x speedup.
\subsection{Kernel 1}
Although the choice of hashmap or CSR was not a bad decision, but there can be a lot of improvement which can be expected. One direction would be to possibly implement CSR using a parallel sort and thread-safe arrays(in-order to optimise the storage), keeping in mind that the CSR implementation also increased the execution time of kronecker.
\newline
One thing which can be pointed out is that kernel 2 and kernel 3 speedup is no way largely affected by the Hashmap or CSR graph representation. Usage of other structures might cause inefficiency but as far as CSR and Hashmaps are concerned, both have been tested out for kernel2 and gives a similar metric for speedup.
\subsection{Extension of Metrics to TEPS : Using a bigger machine than 24 cores}
Graph500 expects the performance metric to be edges traversed in a second and therefore, a major amount of testing is required. One additional point to remember is that kernel 2 and 3 have implemented only for a single start vertex. Graph500 expects the metric to be averaged out over 64 such start vertices. Because of the machine limitation, we preferred speedup metric because the set we worked on is a much smaller graph size than what graph500 expects.    

\bibliographystyle{unsrt}
\bibliography{main}

\end{document}